\documentstyle[preprint,prc,aps,epsfig]{revtex}
\begin{document}

\draft
\tightenlines
\title{Low-energy p-d Scattering: High Precision Data, Comparisons
with Theory, and Phase-Shift Analyses}
\author{M.~H.~Wood\footnote{Present Address: Thomas Jefferson National
Accelerator Facility, 12000 Jefferson Ave., Newport News, VA, 23606},
C.~R.~Brune, B.~M.~Fisher, H.~J.~Karwowski, D.~S.~Leonard, and E.~J.~Ludwig}
\address{Department of Physics and Astronomy, University of North
Carolina at Chapel Hill\\
Chapel Hill, NC 27599-3255, USA\\
and Triangle Universities Nuclear Laboratory, Durham,
NC  27088-0308, USA}
\author{A.~Kievsky, S.~Rosati, and M.~Viviani} 
\address{Istituto Nazionale di Fisica Nucleare, Sezione di Pisa and
Dipartimento di Pisa, Universita di Pisa, I-56100 Pisa, Italy} 

\date{\today}
\maketitle
\widetext

\begin{abstract}
Angular distributions of $\sigma(\theta)$, $A_{y}$, $iT_{11}$,
$T_{20}$, $T_{21}$, and $T_{22}$ have been measured for d-p scattering
at $E_{c.m.}=667$~keV.  This set of high-precision data is
compared to variational calculations with the nucleon-nucleon
potential alone and also to calculations including a three-nucleon (3N)
potential. Agreement with cross-section and tensor analyzing
power data is excellent when a 3N potential is
used. However, a comparison between the vector analyzing powers
reveals differences of approximately 40\% in the maxima of the
angular distributions which is larger than reported at higher energies
for both p-d and n-d scattering. Single-energy phase-shift analyses were
performed on this data set and a similar data set at
$E_{c.m.}=431.3$~keV.  The role of the different phase-shift parameters
in fitting these data is discussed. 
\end{abstract}


%
\section{Introduction}
\label{sec:intro}
An important question in low-energy nuclear physics is whether a
nucleon-nucleon (NN) interaction is sufficient to describe few-nucleon 
systems or whether a three-nucleon force (3NF) is necessary when the
system consists of A$\geq 3$.  To solve for the bound and scattering
states of the three-nucleon (3N) system, state-of-the-art theoretical 
methods like the Faddeev method~\cite{Glockle83} (which works well for
n-d scattering) have been developed.  It was not until the advancement  
of variational techniques~\cite{Kievsky94} which could include the
Coulomb force rigorously that serious comparisons could be made with p-d
scattering data.  Current theoretical investigations incorporate
phenomenological NN potentials such as CD Bonn~\cite{Jiang92},
Nijmegen III~\cite{Stoks94}, and Argonne $v_{18}$
(AV18)~\cite{Wiringa95}.  These potentials are based on the
one-pion-exchange potential for large interparticle separations as
well as fits to the present NN database. When calculations using those
models are performed in the three-nucleon system, the ${\rm ^{3}He}$ and
${\rm ^{3}H}$ binding energies are underestimated by approximately
$0.5$ to $1.0$~MeV depending on which NN potential is employed. The
under-prediction of the $A=3$ binding energies is the first indication
that the two-nucleon interaction alone, as parameterized by the NN
potentials in their current forms, is insufficient for an accurate
description of the 3N data. To remedy the situation, phenomenological
3N potentials such as the Tuscon-Melbourne (TM)~\cite{Coon79}, the
Brazil (BR)~\cite{Coelho83}, and the Urbana IX
(UR-IX)~\cite{Pudliner95} were introduced which are based on the
2$\pi$ exchange interaction and adjusted to reproduce the triton
binding energies.  By applying chiral-perturbation theory to the 3NFs,
Friar {\it et al.}~\cite{Friar99} showed that the present
phenomenological 3NFs are equivalent to first-order in the Lagrangian
they constructed except for one term in the TM potential which was
found to be of next-to-leading order.

\begin{figure}
\centering
\epsfig{file=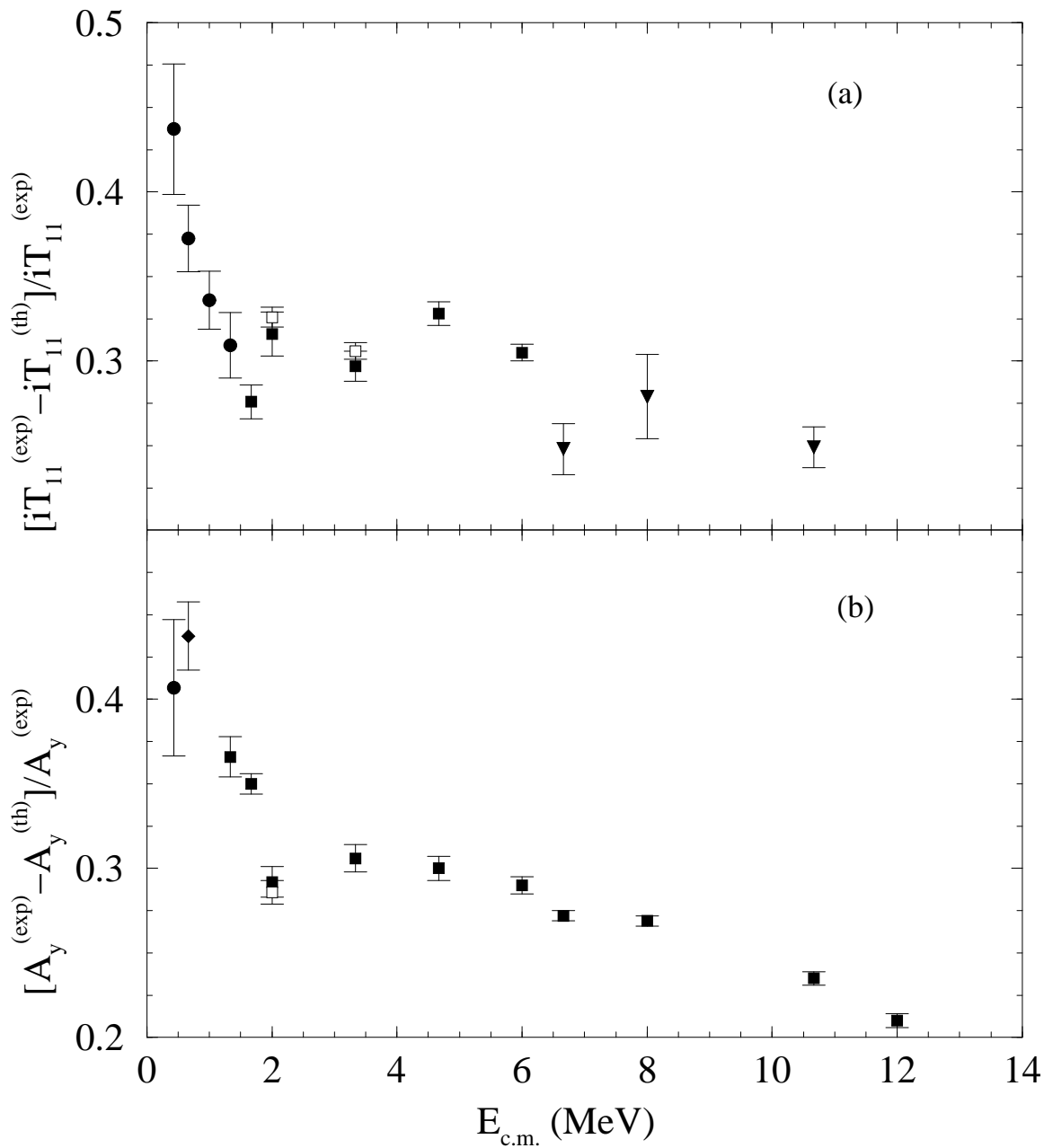,width=6.0in}
\caption{Differences between experimental data and theoretical
  calculations using the AV18 potential at the maximum of the angular
  distributions for (a) $iT_{11}$ and (b) $A_{y}$.  The points are
  plotted as the difference between experiment and theory divided by the
  experimental value.  The data are taken from the following sources:
  Ref.~\protect\cite{Brune00,Karwowski99} (circles),
  Ref.~\protect\cite{Shimizu95,Sagara94,Sagara00} (filled squares),
  Ref.~\protect\cite{Knutson93,Sowinski87} (open squares),
  Ref.~\protect\cite{Gruebler83} (triangles),
  and Ref.~\protect\cite{Huttel83} (diamond).}
\label{fig:aypuzzle}
\end{figure}

Beyond the binding-energy problem, there is another large discrepancy
in the properties of the 3N system, namely the vector analyzing powers
(VAPs) $A_{y}$ and $iT_{11}$ for N-d scattering.  Comparisons between
variational calculations~\cite{Kievsky96} including one of the NN
potentials and VAP measurements at $E_{c.m.} \leq
2$~MeV~\cite{Knutson93,Shimizu95} show that the calculations
under-predict the data by $\approx 30$\% in the maximum of the angular
distributions.  The disagreement becomes worse at 
lower energies where the difference increases to $\approx 40$\% at
$E_{c.m.}=431.3$~keV~\cite{Brune98,Brune00}.  This VAP problem was
first observed for $A_{y}$ in n-d scattering and has been labelled the
``$A_{y}$ puzzle''~\cite{Tornow91,Witala91}.  The trend of the ``$A_{y}$
puzzle'' for both $A_{y}$ and $iT_{11}$ is a decreasing discrepancy as
the energy increases as shown in Fig.~\ref{fig:aypuzzle}.  The
disagreement mostly disappears at   $E_{c.m.} \approx
30$~MeV~\cite{Glockle96}.  The inclusion of a 3NF in the variational
calculations lessens the discrepancies with $A_{y}$ and $iT_{11}$ data
for p-d scattering at the aforementioned energies by no more than $15$\%. 
The effect of the magnetic moments was investigated by
Stoks~\cite{Stoks98} and found not to provide the solution. 

To investigate this question further, we have measured with high
precision $A_{y}$ and  $iT_{11}$ as well as the cross section
$\sigma(\theta)$ and the tensor  analyzing powers (TAPs) $T_{20}$,
$T_{21}$, $T_{22}$ for p-d elastic scattering at $E_{c.m.}=667$~keV.
To isolate the deficient phase shifts and mixing parameters in the
theoretical calculations, two single-energy phase-shift analyses are
performed, one with these data at $E_{c.m.}=667$~keV and another with
the same observables at
$E_{c.m.}=431.3$~keV~\cite{Brune98,Brune00,Kievsky97,Karwowski99}
previously measured by our group.  Additional information concerning the
experimental details and analysis is available in Ref.~\cite{Wood00}.
Moreover, the 667~keV cross-section measurement was included in a
$\chi^{2}$ study~\cite{Kievsky00} with other cross-section data below
$E_{c.m.}=2$~MeV.
\section{Experimental Details}
\label{sec:expt}
The basic experimental setup and techniques were developed for our
measurements of d-p elastic scattering at $E_{c.m.}=431.3$~keV and are
described in
Ref.~\cite{Brune98,Brune00,Kievsky97,Karwowski99,Wood00,Kievsky00}.
The modifications to the setup described in Ref.~\cite{Brune00} for
the experiments conducted at $E_{c.m.}=667\pm1$~keV will be discussed.
One advantage of making measurements at $E_{c.m.}=667$~keV over 
431.3~keV is the reduction of the Rutherford scattering of the
incident particles from carbon which was present in the targets. The
statistical uncertainties in this set of measurements are
significantly smaller than in the measurement at 431.3~keV.  The only
slight disadvantage is the presence of additional proton
groups from the ${}^{12}{\rm C}(d,p){}^{13}{\rm C}$ reaction in the
deuteron beam experiments. All six angular
distribution measurements employed the FN tandem accelerator at the
Triangle Universities Nuclear Laboratory to give a deuteron
(proton) beam energy in the center of the target of 2.00~MeV (1.00~MeV). 
\subsection{Cross-Section Measurements}
\label{sec:csexpt}
\begin{figure}
\centering
\epsfig{file=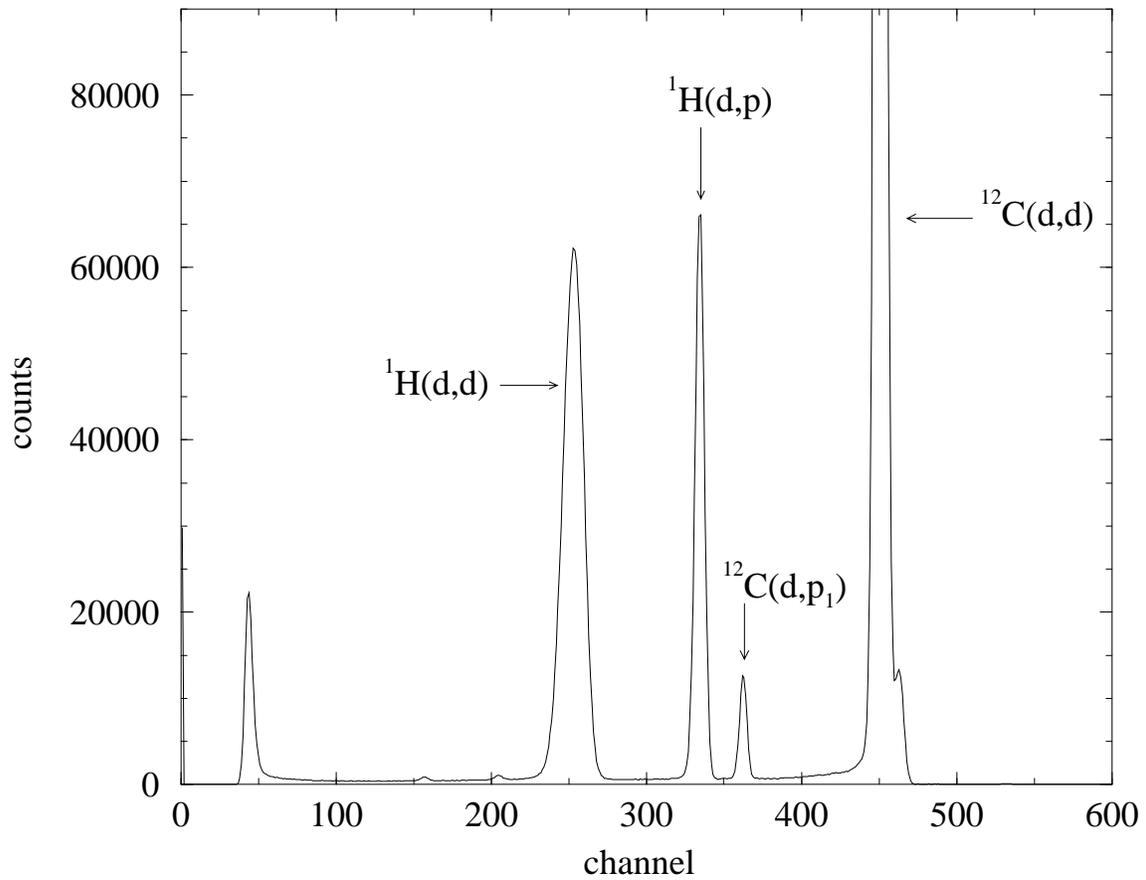,width=6.0in}
\caption{Typical spectrum for a 2.00~MeV deuteron beam on thin
  hydrogenated carbon foil.  The detectors were positioned at 
$\theta_{lab}=26.1^{\circ}$.  In addition to the peak at channel 360,
there are two smaller peaks at channels 160 and 210 which arise from
$^{12}{\rm C}(d,p)^{13}{\rm C}$ reactions.}  
\label{fig:csspec}
\end{figure}

\begin{figure}
\centering
\epsfig{file=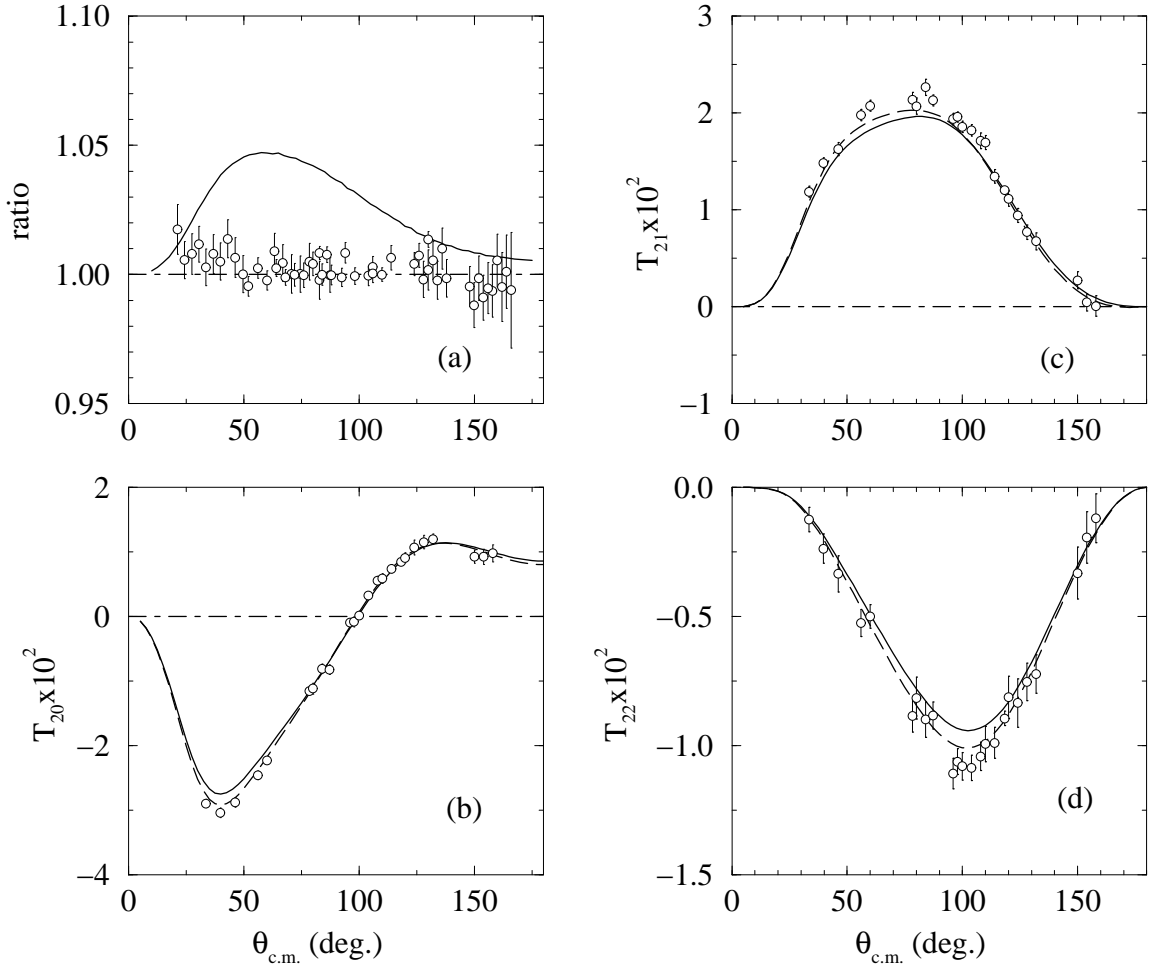,angle=270,width=6.0in}
\caption{Angular distributions for p-d elastic scattering at
    $E_{c.m.}=667$~keV. Graph (a) shows the ratio of the cross-sections 
    to the AV18+UR-IX calculation.  The circles represent the data, including
    statistical errors, divided by the AV18+UR-IX calculation. The
    solid line is the ratio of the AV18 calculation to the AV18+UR-IX
    calculation.  Graphs (b), (c), and (d) show the angular
    distributions of the TAP data.  The solid and dashed curves are
    calculations with the AV18 and AV18+UR-IX potentials,
    respectively.  The dot-dashed line marks zero on these graphs.}  
\label{fig:taps}
\end{figure}

The relative cross-section was measured with a beam of deuterons
on thin hydrogenated carbon targets~\cite{Black95} which contained
approximately $(0.5\pm0.1)\times10^{18}~\rm{H/cm^{2}}$ and
$(1.0\pm0.2)\times10^{18}~\rm{C/cm^{2}}$. The rest of the experimental
details can be found in Ref.~\cite{Kievsky00}.  A sample spectrum for 
${}^1{\rm H}(d,d){}^1{\rm H}$ scattering at
$\theta_{lab}=26.1^{\circ}$ is shown in Fig.~\ref{fig:csspec}.  The
absolute cross-section was obtained by normalizing the relative
$\sigma(\theta)$ for d-p scattering to the well-known  
${}^1{\rm H}(p,p){}^1{\rm H}$ cross-section determined by the Nijmegen
phase-shift analysis~\cite{Nim}. The uncertainties of the absolute
cross-section measurement are 0.5\% from counting statistics, 0.5\%
from the normalization to the p-p scattering cross-section, and 0.6\%
from the beam-current integration.  The absolute $\sigma(\theta)$
angular distributions are shown in Fig.~\ref{fig:taps}a.  The data are
divided by the results of the variational calculation to emphasize the
differences between the data and the calculations.
\subsection{Analyzing Power Measurements}
\label{sec:apexpt}
For the five analyzing power measurements, the techniques used were
similar to those described in Refs.~\cite{Brune98,Brune00}. The
targets~\cite{Black95} contained 
$(1.0\pm0.2)\times10^{18}$ hydrogen or deuterium atoms and
$(2.0\pm0.4)\times10^{18}$ carbon atoms per ${\rm cm^{2}}$.  The beam
polarization for each experiment was measured online with a
polarimeter situated behind the scattering chamber. For the $T_{20}$,
$T_{21}$, and $T_{22}$ experiments, a polarimeter~\cite{Tonsfeldt80}
based on the ${}^3{\rm He}(\vec{d},p){}^4{\rm He}$ reaction was
utilized.  The beam polarizations for the $T_{20}$ and
$T_{22}$ measurements were $p_{zz} \approx \pm 0.8$ and $p_{zz} \approx
0$ for the three spin states with a 3\% error.  For the $T_{21}$
measurement, the beam polarizations were $p_{zz} \approx \pm
0.8$ and $p_{zz} \approx 0$ with a 2\% error.  For $iT_{11}$, a
polarimeter~\cite{Wood00} based on the ${}^2{\rm H}(\vec{d},p){}^3{\rm
H}$ reaction was used, and the beam polarizations were  $p_{z}
\approx \pm 0.55$. For the proton beam, the 
${}^6{\rm Li}(\vec{p},{}^3{\rm He}){}^4{\rm He}$
reaction~\cite{Wood00,Brune97} was employed for polarimetry.  The beam
polarization over the duration of the experiment was $p_{z} \approx
\pm 0.65$. The error in the beam polarizations for both experiments was 2\%.

\begin{figure}
\centering
\epsfig{file=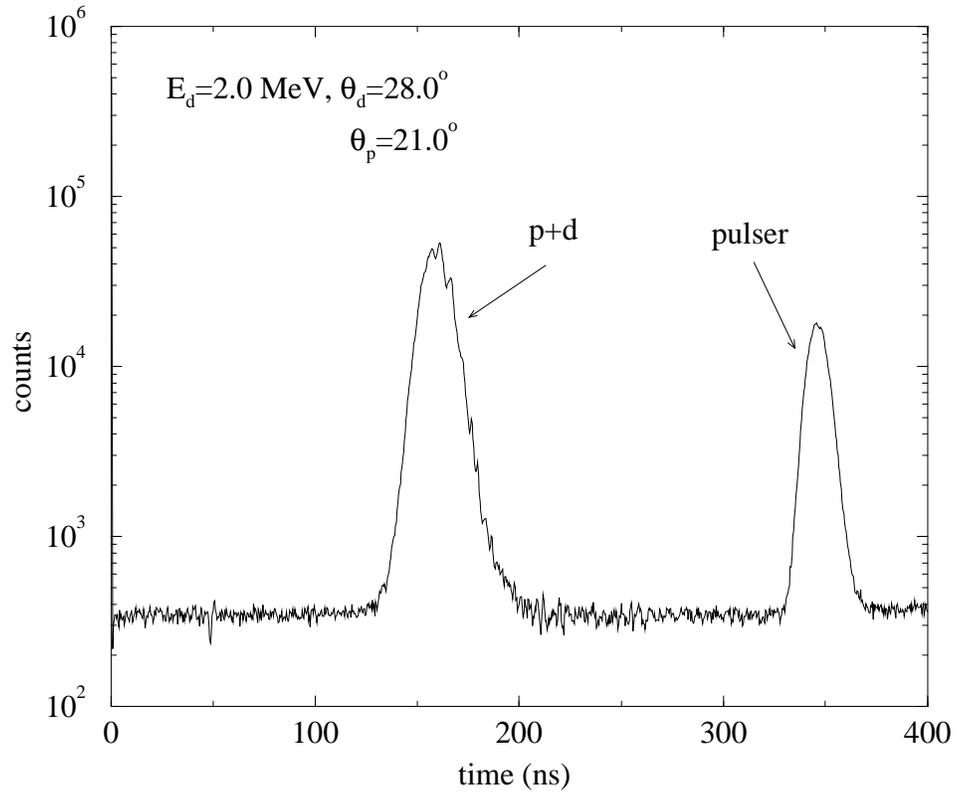,width=5.0in}
\caption{A proton-dueteron time-of-flight spectrum using a four-detector
  configuration.  The time resolution is approximately 20~ns.}
\label{fig:coinspec}
\end{figure}

\begin{figure}
\centering
\epsfig{file=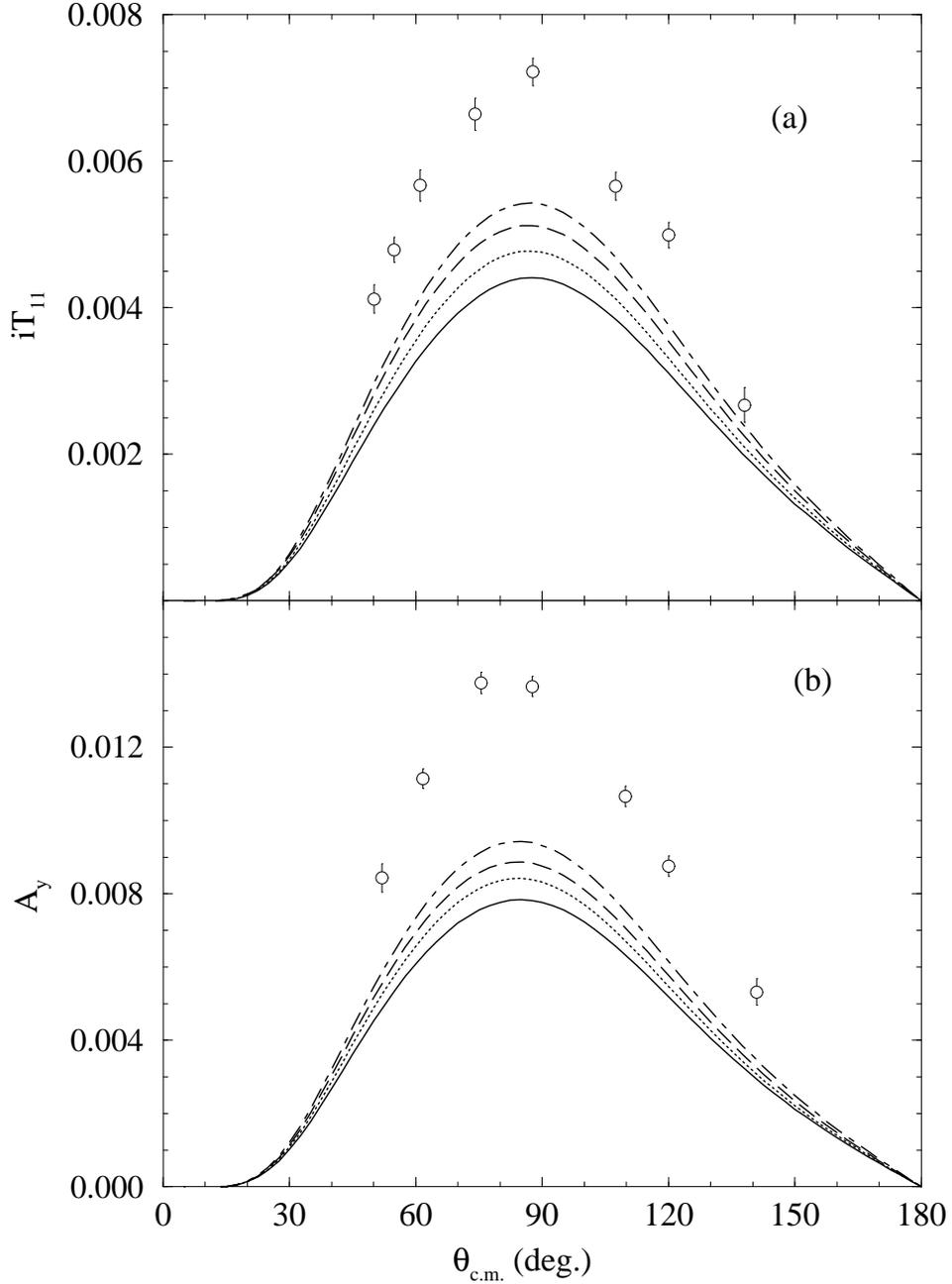,width=5.0in}
\caption{Angular distributions for $iT_{11}$ and $A_{y}$ for p-d
  elastic scattering at $E_{c.m.}=667$~keV.    The errors include the
  uncertainty in the beam polarization as well as statistical
  uncertainties.  The solid, dashed, and dotted curves are variational
  calculations with the AV18, AV18+UR-IX, and AV18+TM potentials,
  respectively.  The dot-dashed curve is the calculation with the
  AV18+TM$^{\prime}$ potentials.}
\label{fig:vaps}
\end{figure}

The coincidence technique described in Ref.~\cite{Brune98} was used
for the VAP experiments.  A new addition was 2-${\rm \mu m}$-thick 
mylar foils that were placed in front of the detectors to
block any heavy recoil particles such as carbon atoms. By making the
measurements at the higher energy, it was possible to cover a larger
angular range ($\theta_{c.m.}\approx50^{\circ}-140^{\circ}$) than obtained at
431.3~keV.  A typical timing spectrum for p-d coincidences for a
deuteron beam incident on a hydrogen target is shown in
Fig.~\ref{fig:coinspec}. The TAP and VAP angular distributions are
given in Figs.~\ref{fig:taps} and~\ref{fig:vaps}, respectively. 

\section{Comparisons with the Theoretical Calculations}
Theoretical calculations to compare with the data have been produced
using the AV18 potential and the AV18 plus a 3NF. These calculations
employ the Pair-Correlated Hyperspherical Harmonic basis to construct
the scattering wave functions and the Kohn variational principle to
solve for the scattering-matrix elements~\cite{Kievsky94}.  In order to study
the sensitivity to different 3NF models, the UR-IX, the TM and its
modified version denoted TM$^{\prime}$ interactions have been
considered. Originally the TM potential~\cite{Coon79} was given as a sum of
four terms characterized by the strengths $a,b,c,d$. To assure consistency
with chiral symmetry the $a$-term in the TM model has been redefined
with a new constant $a'=a-2m^2_\pi c$ (which essentially implies a
change of sign for the constant $a$) and the $c$-term has been set to
zero~\cite{Friar99}. The TM$^{\prime}$ model, as well as the TM and the
UR-IX potentials, have been used recently in an extensive study of 3NF
effects in n-d scattering above $E_{c.m.}=2$ MeV~\cite{Witala00}. For
completeness, here we will present results using the same set of
constants as Ref.~\cite{Witala00} for the TM and TM$^{\prime}$ models.

Accompanying the data presented in Figs.~\ref{fig:taps} and~\ref{fig:vaps}
are several separate variational calculations, one including the AV18
potential and the others with the AV18 and either the UR-IX or TM
potentials. It is clear that the calculations with or without the 3N
potential reproduce $\sigma(\theta)$ and the TAP data fairly well.
On the other hand, the agreement with the VAP data is much poorer.
These results are qualitatively similar to those observed 
at higher energies which were mentioned in the Introduction.  For the
present case, the differences between the NN-potential-only calculations
and the $A_{y}$ and $iT_{11}$ measurements are approximately 40\% and
37\%, respectively at the maximum of the angular distributions.  With 
the inclusion of the UR-IX potential, the agreement for all of the
observables improves; however, the change in the VAPs is only marginal
and a large discrepancy still remains on the order of 36\% and 29\% for $A_{y}$ and $iT_{11}$, respectively.  When the TM 3N-potential is
included, the  VAP calculations show slightly worse agreement with the
data (39\% and 34\% differences) than in the case of the UR-IX
potential. However, when the TM$^{\prime}$ potential is used, the
calculations are closer to the VAP data than for the case of the UR-IX
potential; the difference between $A_{y}$($iT_{11}$) data and the
calculations is 32\%(25\%). The same trend has been observed in
Ref.~\cite{Witala00} at $E_{c.m.}=2$ MeV. The difference between the
calculations with the TM$^{\prime}$ potential and those with UR-IX is
about 5\%. While each of the 3NFs mentioned here provide some
improvement to the $A_{y}$ and $iT_{11}$ calculations, none of the
potentials eliminates the descrepancy. To obtain a more quantitative
comparison, the reduced $\chi^{2}$, $\chi_{N}^{2}$, was calculated for
each observable, and the results are described in Sec.~\ref{sec:chi2}. 

\begin{figure}
\centering
\epsfig{file=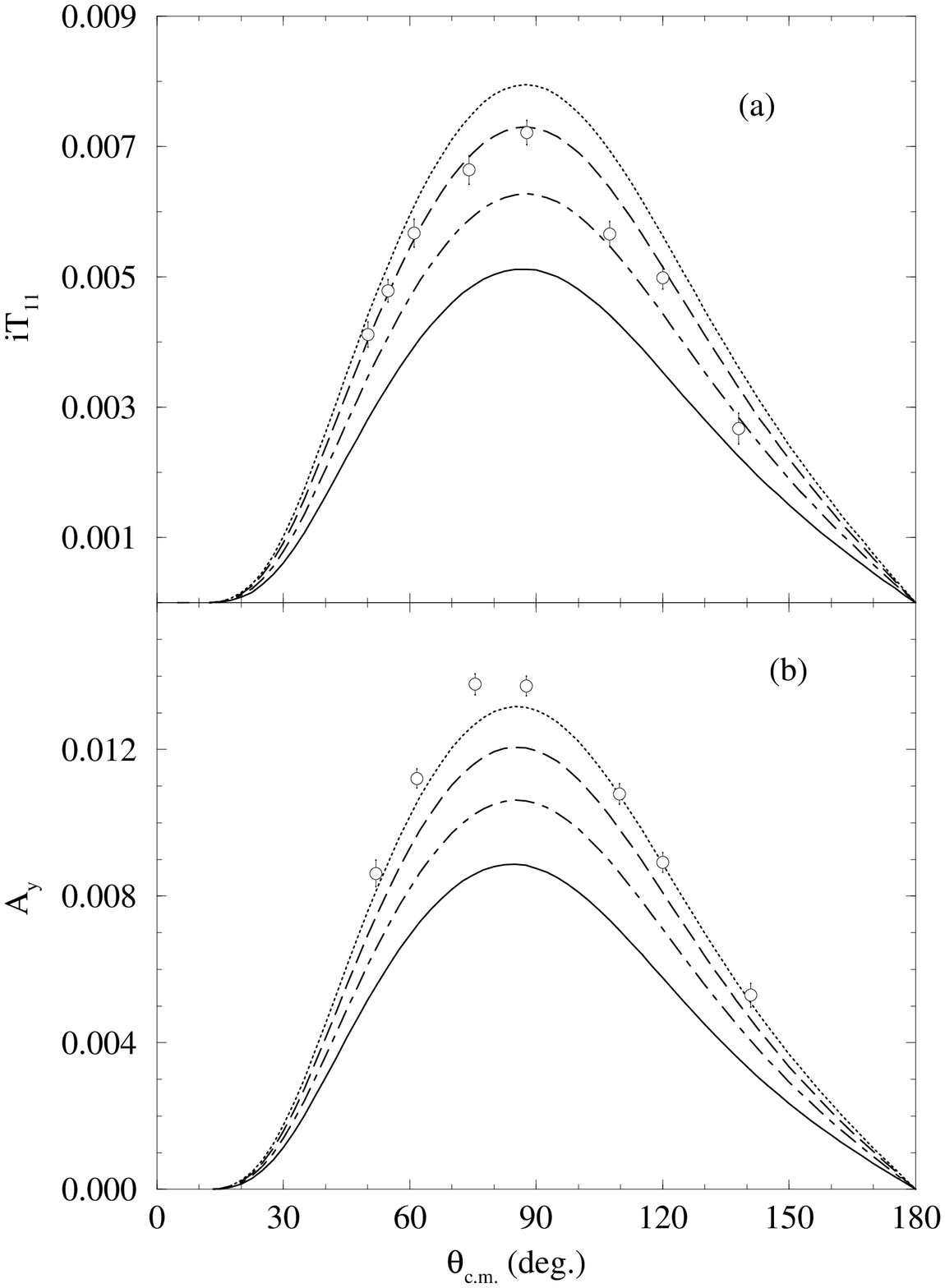,width=5.0in}
\caption{Comparison between $iT_{11}$ and $A_{y}$ data and calculations with
AV18+UR-IX with the addition of phenomenological spin-orbit forces at
$E_{c.m.}=667$~keV. The solid curve is the calculation without any
modifications.  The dotted curve represents a long-range interaction
($W_{0}=-1$~MeV, $\alpha=0.7~{\rm fm}^{-1}$), the dashed curve is for a
medium-range ($W_{0}=-10$~MeV, $\alpha=1.2~{\rm fm}^{-1}$), and the
dot-dashed curve indicates a short-range interaction ($W_{0}=-20$~MeV,
$\alpha=1.5~{\rm fm}^{-1}$)~\protect\cite{Kievsky99}.} 
\label{fig:lsvaps}
\end{figure}

Besides the TM$^{\prime}$ potential, other modified NN and 3N force
models have recently become available. For example, a new NN interaction
constructed from chiral perturbation theory seems to give a better
description of the VAP data at low energy~\cite{Epelbaum01}. A different
possibility may be in the construction of a new 3N force.  H\"uber and
Friar~\cite{Huber98} have suggested that a possible candidate should
be a spin-dependent 3NF since the VAP is a difference between
polarized cross-sections.  Following this direction, a phenomenological 
3N $\vec{L}\cdot\vec{S}$ force has recently been proposed~\cite{Kievsky99}.  
This new force essentially modifies the scalar function $v^{ls}_{11}(r)$
already present in the $\vec{L}\cdot\vec{S}$ term of the AV18 potential
in triplet spin and isospin channels. Explicitly the 
following form has been proposed 
\begin{equation}
   V^{ls}_{3N}= \sum_{i<j}v^{ls}_{11}(r_{ij})
          {\bf L}_{ij}\cdot{\bf S}_{ij}P_{11}(ij)+
          W_0{\rm e}^{-\alpha\rho}\sum_{i<j}
          {\bf L}_{ij}\cdot{\bf S}_{ij}P_{11}(ij) \ .
\label{eq:ls3}
\end{equation}
where $P_{11}$ is a projector in channels with spin $S_{ij}=1$ and
isospin $T_{ij}=1$. The hyperradius $\rho$ is
\begin{equation}
   \rho = \sqrt{{2\over 3}(r_{12}^2+r_{23}^2+r_{31}^2)}
\end{equation}
and $W_0$ and $\alpha$ are parameters characterizing the strength and range
of the three-body term. This force influences the splitting of the
$^{4}P_{J}$ phases and the magnitude of $\epsilon_{{3/2}^{-}}$ without
appreciably affecting the other parameters.  Moreover, when the deuteron and
spectator nucleon are separated by a large distance, the proposed interaction
reduces to the original NN $\vec{L}\cdot\vec{S}$ potential
$v^{ls}_{11}(r_{ij})$. Fig.~\ref{fig:lsvaps} shows $A_y$ and $iT_{11}$
calculated using this new potential for three different strength and
range parameters.  The values for $W_0$ and $\alpha$ have been taken
from Ref.~\cite{Kievsky99} and have been fixed in order to reproduce
the VAP data at $E_{c.m.}=2$ MeV. The addition of the 3N spin-orbit force
has provided much better agreement with our VAP data than previous
calculations. The various phase-shift parameters will be discussed in
detail in Sec.~\ref{sec:psa}.  Moreover the  calculated differential
cross sections and TAPs remain essentially unchanged~\cite{Kievsky99}
by the new term. 

\section{Calculations of Reduced $\chi^{2}$}
\label{sec:chi2}
To investigate quantitatively the effects of the UR-IX potential,
reduced $\chi^{2}$ calculations were completed with
both the AV18 potential and the AV18+UR-IX potentials.  More details
for the $\sigma(\theta)$ comparison are presented in
Ref.~\cite{Kievsky00}, and the results are listed in the first two
lines of Table~\ref{ta:psatrials667}.  None of the data were
renormalized in order to find a minimal $\chi_{N}^{2}$. From
$\chi_{N}^{2}$ calculations, it is possible to draw conclusions
similar to those from the visual comparisons: the $\sigma(\theta)$ and
the TAP data are reproduced very well while VAP data show 
poorer agreement with theory.  However, all of
the $\chi_{N}^{2}$ are improved when a 3NF is included.  For
$\sigma(\theta)$, $\chi_{N}^{2}$ improves by an order of
magnitude. The $\chi_{N}^{2}$ is reduced significantly 
for each TAP.  The calculations of $A_{y}$ and $iT_{11}$ with UR-IX show
improved agreement with the data; however, the $\chi_{N}^{2}$ remains in
the hundreds.
\section{Phase-Shift Analysis}
\label{sec:psa}
The technique of Phase Shift Analysis (PSA) is useful for determining the
dependences of the observables on the various partial waves, and also
for studying the sources of disagreements between theory and
experiment. Here we consider data at two energies: $E_{c.m.}=667$~keV, the
energy of the present data, and also $E_{c.m.}=431.3$~keV, the energy of
previous measurements in our laboratory~\cite{Brune00}. The PSAs
reported here are carried out in the framework of
Seyler~\cite{Seyler69}, where the scattering matrix is parameterized
in terms of phase shifts and mixing parameters. The phase shift parameters in
general depend on orbital angular momentum $L$, total angular momentum
$J$, and the channel spin $S$. The mixing parameters connect partial
waves with different $L$ and/or $S$ for a given $J$ and parity.

Before embarking upon a detailed PSA, it is important to understand the
convergence of the observables versus the angular momentum cutoff.
Studies were performed using the AV18+UR-IX calculations for the
phase shifts and mixing parameters. The dependence on the angular
momentum cutoff $L_{max}$ was examined by setting to zero all phase
shifts and mixing parameters with $L>L_{max}$ for different values of
$L_{max}$. Note that the effect of the Coulomb potential is retained in all
partial waves, so it is expected that the observables will converge quickly.
We have quantified the convergence of the observables by using
$\chi^2_N$ comparisons of the calculations with different $L_{max}$ values
to the experimental data at $E_{c.m.}=431.3$ and 667~keV.
The results are shown in Table~\ref{tab:conv}.
First of all we find very little change in $\chi^2_N$
between $L_{max}=3$ and $L_{max}=4$, indicating, as expected, that the
results are already well-converged with $L_{max}=3$ at these low energies.
It is also interesting to note that the cross section and TAPs
are fairly well described with $L_{max}=2$, while only $L_{max}=1$
is required to produce a similar level of convergence for the VAPs.
Smaller, but statistically-significant improvements are obtained
by extending to $L_{max}=3$. We have also observed in these studies that
the observables $T_{20}$ and $T_{22}$ are highly dependent on the
$L=2$ phase shifts and mixing parameters.

Single-energy PSAs have been performed by using the phase shifts and
mixing parameters produced by the AV18+UR-IX potential as starting
values. Some of the parameters are allowed to vary freely; the
best-fit values for the variable parameters are determined by
$\chi^{2}$ minimization with attention given to the fitting procedure to avoid
local minima. Phase shifts were included up to a maximum
angular momentum of $L=4$, but only those with $L \leq 2$ were allowed to vary.

Previous phase-shift analyses~\cite{Kievsky96,Knutson93} have linked
the VAP problem to the $P$-wave phase-shifts.  To test the effects of
the different $P$-wave phase-shifts and their associated mixing
parameters, two-parameter fits where made. 
Tables~\ref{ta:psatrials667} and~\ref{ta:psatrials431} show the
$\chi^{2}_{N}$ at the two energies for sample trials where only 
the $P$-wave phase-shifts were varied.  By varying the $^{2}P_{J}$
phase-shifts and/or the mixing parameter $\epsilon_{{1/2}^{-}}$
freely (trials 1 and 2), there is very little change in the $\chi^{2}_{N}$
with the exception of $iT_{11}$ at 431.3~keV.  With the
combination of $\epsilon_{{3/2}^{-}}$ and a $^{4}P_{J}$
phase-shift (trials 3, 4, and 5), good fits were produced with the most
dramatic improvement occurring in each VAP.  At $E_{c.m.}=667$~keV, the
$\chi^{2}_{N}$ reduced by a factor of approximately 300 for $A_{y}$
and a factor of 30 for $iT_{11}$.  For the $E_{c.m.}=431.3$~keV case,
the fits produced reductions in $\chi^{2}_{N}$ by factors of
approximately 20 and 10 for $A_{y}$ and $iT_{11}$, respectively, with
the exception of the fit to $iT_{11}$ with the free parameters
$^{4}P_{3/2}$ and $\epsilon_{{3/2}^{-}}$.  A comparison of the three
$^{4}P_{J}$ trials indicates that the two-parameter fit with the
$^{4}P_{1/2}$ phase-shift and $\epsilon_{{3/2}^{-}}$ gives the best
result although the differences with the other two trials are very
small. These trials so far illustrate the influence of the $P$-wave
phase-shifts and $\epsilon_{{3/2}^{-}}$; however, a more realistic fit 
(trial 6) is to allow all of the $S$- and $P$-wave phase-shifts with
$\epsilon_{{1/2}^{-}}$ and $\epsilon_{{3/2}^{-}}$ to vary freely (9
parameters in all).  The remaining phases and mixing parameters were
taken from the AV18+UR-IX calculations~\cite{Kievsky96}. The results
of the fits are listed in Table~\ref{ta:psa}.

While there were improvements with the full $S$- and $P$-wave fits
over the two parameter $^{4}P_{J}$ fits, the reduction in
$\chi^{2}_{N}$ is small.  To make a more quantitative comparison
between trials, errors were determined for specific
parameters by the change in a specific parameter necessary to increase the
minimal $\chi^{2}$ by 1~\cite{Bevington92}. Allowing more parameters to
vary led to a larger error in each parameter which underlies our attempts
to find the best fit with the least number of parameters.  The
uncertainties in phase shifts and mixing parameters when all $S$- and
$P$-wave parameters were varied are given in Table~\ref{ta:psa}. These
errors reflect the contribution of statistical uncertainties in the
data only.  

The PSA presented here produced very good fits to the data sets at
$E_{c.m.}=431.3$~keV and 667~keV with sensible results.  
The greatest influence on the $A_{y}$ and $iT_{11}$ fits came from the
$^{4}P_{J}$ phase-shifts and $\epsilon_{{3/2}^{-}}$ parameter.
This result was noticed in the earlier analyses~\cite{Kievsky96,Knutson93}
at higher energies.  It is difficult to provide a more precise value
from the PSA for each phase-shift with the present data sets at
431.3~keV and 667~keV.  What these analyses do confirm is the need to
modify the $^{4}P_{J}$ phase-shifts and the mixing between
the $^{4}P_{3/2}$ and the $^{2}P_{3/2}$
phase-shifts (as evident from the change in $\epsilon_{{3/2}^{-}}$).
In fact, $\epsilon_{{3/2}^{-}}$ was increased (in absolute value) by
about $19\%$ from the 667~keV PSA and by about $28\%$ from the 431.3~keV PSA
over the predicted value even when the UR-IX potential was included in
the calculations.  The reader should be cautioned that the errors from
both analyses allow much wider ranges of values for $\epsilon_{{3/2}^{-}}$.
The theoretical calculations also appear to
underestimate the splitting $\Delta P= {}^4P_{5/2}- {}^4P_{1/2}$.  
For example at 667 keV the calculated splitting is $\Delta P=0.87^\circ$
compared to $0.96^\circ$ obtained in the fit. At 431.3 keV
the difference is even bigger; the calculated value is 
$\Delta P=0.48^\circ$ compared to $0.63^\circ$ from the fit.
The small, but nevertheless important, underestimation of 
$\Delta P$ has been found also in the calculations at $E_{c.m.}=2$
MeV~\cite{Kievsky96}. 

\section{Conclusions}
\label{sec:conclusions}
In conclusion, we have presented high-precision measurements of
$\sigma(\theta)$, $A_{y}$, $iT_{11}$, $T_{20}$, $T_{21}$, and $T_{22}$ 
for p-d elastic scattering at a center-of-mass energy of $667$~keV.
The cross section and TAP data show excellent agreement with the
variational calculations using the AV18 and UR-IX potentials.  The
``$A_{y}$ puzzle'' persists at this energy with the inclusion of the
UR-IX in the calculations giving a small improvement in the agreement
with the VAP data.  In order to extend the analysis to different forms
of the 3NFs, the AV18+TM and AV18+TM$^{\prime}$ potential models have also been
considered in the comparisons to the data.  The calculations with
either TM potential under-predict the VAP data by a similar amount
as in the case of the AV18+UR-IX results.  From the reduced $\chi^{2}$
analysis, it is clear that adding a 3NF into the calculations improves
the agreement with the data for all of the observables. 

From a single-energy PSA at very low energies ($E_{c.m.}=431.3$~keV
and 667~keV), the under-prediction of the VAP data by the
calculations appears to be caused by certain $P$-wave components.  By
varying two parameters, $\epsilon_{{3/2}^{-}}$ and one of the
$^{4}P_{J}$ phase-shifts, and maintaining all of the other parameters as
determined from theory, an almost perfect fit was achieved for
$A_{y}$.  For $iT_{11}$, the $\chi_{N}^{2}$ was reduced by more than an
order of magnitude to approximately 1.3 for the 431.3~keV PSA and 1.8
for the 667~keV analysis.  To obtain an even better
fit, all of the $S$- and $P$-wave phase-shifts along with
$\epsilon_{{1/2}^{-}}$ and $\epsilon_{{3/2}^{-}}$ were varied freely 
to obtain $\chi^{2}_{N}$ of $\approx1.0$ for each observable.  The
results of this analysis are the same within the errors as the PSA with
$\epsilon_{{3/2}^{-}}$ and a $^{4}P_{J}$ phase-shift as the only
free parameters.  From comparisons of the PSA results to calculations it is
evident that the theory underestimates the splitting of the $^{4}P_{J}$
phase shifts as well as the mixing between $^{2}P_{3/2}$ and
$^{4}P_{3/2}$.

The inclusion of a phenomenological 3N spin-orbit force, which
fixes the $P$-wave mixing in the calculations, provides better
agreement with the VAP data without disturbing the agreement
obtained for the cross-section and TAP data. While this fact does not
determine the existence of a 3NF, it indicates a need to develop and refine the
current 3N potential models~\cite{Shadow01}. In addition, one can not
rule out the possibility that currently employed NN $^{3}P_{J}$
interactions are incorrect~\cite{Epelbaum01}.

Finally, the $A_y$ problem can be studied with information from
p-${\rm ^{3}He}$ elastic scattering data.  The variational
calculations employed in this paper have recently been extended to
the p-${\rm ^{3}He}$ scattering observables~\cite{Viviani01}.  In this
case there is an underprediction by approximately 40\% of new
$A_y$ data taken at an $E_{c.m.}$ below 2~MeV.
Moreover, $A_y$ for p-${\rm ^{3}He}$ scattering is an order of
magnitude larger than for the p-d scattering case.  This fact makes it
an ideal candidate for future investigations, and more p-${\rm
^{3}He}$ elastic scattering data at low energies is needed.

\acknowledgments
The authors would like to thank Dr.~K.~D.Veal for his assistance in
the data taking.  This work was supported in part by the US DOE under
Grant No. DE-FG02-97ER41041. 

\newpage
\begin{table}
\centering
\caption{The $\chi^{2}_{N}$ calculated for each trial of the
  phase-shift analysis for each observable measured at
  $E_{c.m.}=667$~keV. The number of data points in the angular
  distribution of each observable is the value in parentheses under
  the title.  The total $\chi^{2}_{N}$ is the sum of $\chi^{2}$ for
  each observable divided by 143 (the total number of data points).}
\begin{tabular}{c|c|ccccccc}
 & Free & \multicolumn{7}{c}{$\chi^{2}_{N}$} \\
 Trial & Parameters & $\sigma(\theta)$ & $A_{y}$ & $iT_{11}$ & $T_{20}$ &
 $T_{21}$ & $T_{22}$ &  Total \\
 & & (56) & (7) & (8) & (24) & (24) & (24) & (143) \\ \hline
AV18 & None & 45.2 & 275.8 & 112.4 & 3.5 & 3.5 & 2.8 & 39.3 \\ \hline
AV18+UR & None & 1.2 & 190.5 & 61.4 & 1.0 & 2.5 & 0.7 & 13.9 \\ \hline
1 & $^{2}P_{1/2}$, $^{2}P_{3/2}$ & 1.1 & 199.8 & 53.3
& 0.9 & 1.9 & 0.7 & 13.8 \\ \hline
2 & $^{2}P_{1/2}$, $\epsilon_{1/2^{-}}$ & 1.2 & 187.7
& 46.3 & 1.9 & 4.9 & 0.8 & 13.5 \\ \hline
3 & $^{4}P_{1/2}$, $\epsilon_{3/2^{-}}$ & 1.1 & 0.6 & 1.8
& 1.7 & 1.2 & 0.7 & 1.2 \\ \hline 
4 & $^{4}P_{3/2}$, $\epsilon_{3/2^{-}}$ & 1.2 & 0.5 & 5.3
& 2.2 & 5.7 & 0.7 & 2.2 \\ \hline 
5 & $^{4}P_{5/2}$, $\epsilon_{3/2^{-}}$ & 1.3 & 0.5 & 1.9
& 0.8 & 1.8 & 0.7 & 1.2 \\ \hline 
6 & $^{2}S_{1/2}$, $^{4}S_{3/2}$, $^{2}P_{J}$, &
1.0 & 0.8 & 1.1 & 0.8 & 0.9 & 0.5 & 0.9 \\
& $^{4}P_{J}$, $\epsilon_{1/2^{-}}$,
$\epsilon_{3/2^{-}}$ & & & & & & &  \\
\end{tabular}
\label{ta:psatrials667}
\end{table}
\begin{table}
\caption{The convergence of the partial wave expansion is indicated by
$\chi^2_N$ for different values of $L_{max}$ at $E_{c.m.}=431.3$ and 667~keV.
The number of data points for each observable is shown in parentheses.}
\begin{tabular}{cc|ccccccc}
&& \multicolumn{7}{c}{$\chi^2_N$} \\
&$L_{max}$ & $\sigma(\theta)$ & $A_{y}$ & $iT_{11}$ & $T_{20}$ &
 $T_{21}$ & $T_{22}$ &  Total \\ \hline \hline
$E_{c.m.}=431.3$~keV && (22) & (3) & (3) & (24) & (22) & (20) & (94) \\
\cline{3-9}
& 1 & 8.45 & 17.97 & 14.16 & 42.15 & 18.08 & 29.73 & 24.32 \\
& 2 & 2.31 & 17.58 & 13.17 &  1.53 &  5.40 &  1.72 &  3.54 \\
& 3 & 1.87 & 17.62 & 13.13 &  1.91 &  2.61 &  1.55 &  2.85 \\
& 4 & 1.86 & 17.62 & 13.12 &  1.94 &  2.61 &  1.53 &  2.85 \\ \hline \hline
$E_{c.m.}=667$~keV && (56) & (7) & (8) & (24) & (24) & (24) & (143) \\
\cline{3-9}
& 1 & 49.22 & 194.82 & 69.74 & 118.49 & 51.34 & 202.61 & 95.22 \\
& 2 &  1.37 & 189.92 & 61.88 &   2.44 & 10.98 &   3.96 & 16.21 \\
& 3 &  1.11 & 190.43 & 61.38 &   1.11 &  2.83 &   0.66 & 13.96 \\
& 4 &  1.15 & 190.46 & 61.39 &   0.98 &  2.52 &   0.66 & 13.91 \\
\end{tabular}
\label{tab:conv}
\end{table}
\begin{table}
\centering
\caption{The $\chi^{2}_{N}$ calculated for specific trials of the
  phase-shift analysis at $E_{c.m.}=431.3$~keV. The number of data
  points in the angular distribution of each observable is the value
  in parentheses under the title.  The total $\chi^{2}_{N}$ is the sum
  of $\chi^{2}$ for each observable divided by 94 (the total number of
  data points).} 
\begin{tabular}{c|c|ccccccc}
 & Free & \multicolumn{7}{c}{$\chi^{2}_{N}$} \\
 Trial & Parameters & $\sigma(\theta)$ & $A_{y}$ & $iT_{11}$ & $T_{20}$ &
 $T_{21}$ & $T_{22}$ &  Total \\
 & & (22) & (3) & (3) & (24) & (22) & (20) & (94) \\ \hline
AV18 & None & 23.2 & 23.5 & 19.3 & 2.9 & 4.2 & 2.1 & 9.0 \\ \hline
AV18+UR & None & 1.9 & 17.6 & 13.1 & 1.9 & 2.6 & 1.5 & 2.8 \\ \hline
1 & $^{2}P_{1/2}$, $^{2}P_{3/2}$ & 1.6 & 25.1 & 2.7
& 1.3 & 1.1 & 1.6 & 2.2 \\ \hline
2 & $^{2}P_{1/2}$, $\epsilon_{1/2^{-}}$ & 1.6 & 30.1
& 2.2 & 1.2 & 0.9 & 1.6 & 2.2 \\ \hline
3 & $^{4}P_{1/2}$, $\epsilon_{3/2^{-}}$ & 1.7 & 0.9 & 1.3
& 1.2 & 0.9 & 1.5 & 1.3 \\ \hline 
4 & $^{4}P_{3/2}$, $\epsilon_{3/2^{-}}$ & 1.7 & 0.5 & 16.6
& 2.0 & 2.5 & 1.5 & 2.4 \\ \hline 
5 & $^{4}P_{5/2}$, $\epsilon_{3/2^{-}}$ & 2.1 & 1.4 & 3.4
& 1.6 & 1.9 & 1.5 & 1.8 \\ \hline 
6 & $^{2}S_{1/2}$, $^{4}S_{3/2}$, $^{2}P_{J}$, &
1.1 & 0.3 & 0.9 & 1.1 & 0.8 & 1.3 & 1.0 \\
& $^{4}P_{J}$, $\epsilon_{1/2^{-}}$,
$\epsilon_{3/2^{-}}$ & & & & & & &  
\\
\end{tabular}
\label{ta:psatrials431}
\end{table}
\begin{table}
\centering
\caption{Phase shifts and mixing parameters obtained from single-energy
  phase-shift analysis where all $S$- and $P$-wave phase-shifts were 
  varied. The first column for each energy contains the values
  calculated with the AV18 and UR-IX potentials and were the starting
  values for the analyses.  The second column lists the values
  produced from the fits to the data.} 
\begin{tabular}{c|cc|cc}
 & \multicolumn{2}{c}{$E_{c.m.}=431.3$~keV} & 
\multicolumn{2}{c}{$E_{c.m.}=667$~keV} \\
 Parameter & Calculation($^{\circ}$) & Fit ($^{\circ}$) &
 Calculation($^{\circ}$) & Fit ($^{\circ}$) \\ \hline 
 $^{2}S_{1/2}$ & -5.811 & $-7.64\pm0.74$ & -10.00 & $-10.53\pm0.59$ \\
 $^{4}S_{3/2}$ & -28.08 & $-27.48\pm0.25$ & -36.93 & $-36.81\pm0.19$ \\
 $^{2}P_{1/2}$ & -1.993 & $-2.17\pm0.44$ & -3.339 & $-2.91\pm0.15$ \\
 $^{2}P_{3/2}$ & -1.975 & $-1.61\pm0.28$ & -3.303 & $-3.42\pm0.13$ \\
 $\epsilon_{1/2^{-}}$ & 2.499 & $1.82\pm0.46$ & 2.954 &
 $3.33\pm0.13$ \\ 
 $^{4}P_{1/2}$ & 5.262 & $5.06\pm0.11$ & 9.199 & $9.18\pm0.07$ \\
 $^{4}P_{3/2}$ & 6.127 & $6.19\pm0.12$ & 10.69 & $10.62\pm0.05$ \\
 $^{4}P_{5/2}$ & 5.743 & $5.69\pm0.10$ & 10.07 & $10.14\pm0.04$ \\
 $\epsilon_{3/2^{-}}$ & -0.886 & $-1.13\pm0.29$ & -1.060 &
 $-1.25\pm0.06$ \\ 
\end{tabular}
\label{ta:psa}
\end{table}

\begin{references}
\bibitem{Glockle83}W.~Gl\"ockle, {\it The Quantum Mechanical Few-Body
    Problem}, Lecture Notes in Physics Vol. 273, Springer-Verlag,
  Berlin (1983).
\bibitem{Kievsky94}A.~Kievsky, M.~Viviani, and S.~Rosati,
  Nucl. Phys. {\bf A577}, 511 (1994).
\bibitem{Jiang92}M.~F.~Jiang, R.~Machleidt, D.~B.~Stout, and
  T.~T.~S.~Kuo, Phys. Rev. C {\bf46}, 910 (1992).
\bibitem{Stoks94}V.~G.~J.~Stoks, R.~A.~M.~Klomp, C.~P.~F.Terheggen,
  and J.~J.~de~Swart, Phys. Rev. C {\bf 49}, 2950 (1994).
\bibitem{Wiringa95}R.~B.~Wiringa, V.~G.~J.~Stoks, and R.~Schiavilla, 
  Phys. Rev. C {\bf 51}, 38 (1995).
\bibitem{Coon79}S.~A.~Coon, M.~D.~Scadron, P.~C.~McNamee,
  B.~R.~Barrett, D.~W.~E.~Blatt, and B.~H.~J.~McKellar,
  Nucl. Phys. {\bf A317}, 242 (1979).
\bibitem{Coelho83}H.~T.~Coelho, T.~K.~Das, and M.~R.~Robilotta,
 Phys. Rev. C {\bf 28}, 1812 (1983).
\bibitem{Pudliner95}B.~S.~Pudliner, V.~R.~Pandharipande, J.~Carlson,
  and R.~B.~Wiringa, Phys. Rev. Lett. {\bf 74}, 4396 (1995).
\bibitem{Friar99}J.~L.~Friar, D.~H\"uber, and U.~van~Kolck,
  Phys. Rev. C {\bf 59}, 53 (1999).
\bibitem{Kievsky96}A.~Kievsky, S.~Rosati, W.~Tornow, and M.~Viviani,
  Nucl. Phys. {\bf A607}, 402 (1996).
\bibitem{Knutson93}L.~D.~Knutson, L.~O.~Lamm, and J.~E.~McAninch,
  Phys. Rev. Lett. {\bf 71}, 3762 (1993).
\bibitem{Shimizu95}S.~Shimizu, K.~Sagara, H.~Nakamura, K.~Maeda, T.~Miwa,
  N.~Nishimori, S.~Ueno, T.~Nakashima, and S.~Morinobu, Phys. Rev. C
{\bf 52}, 1193 (1995).
\bibitem{Brune98}C.~R.~Brune, W.~H.~Geist, H.~J.~Karwowski,
  E.~J.~Ludwig, K.~D.~Veal, M.~H.~Wood, A.~Kievsky, S.~Rosati, and
  M.~Viviani, Phys. Lett. B {\bf 428}, 13 (1998).
\bibitem{Brune00}C.~R.~Brune, W.~H.~Geist, H.~J.~Karwowski,
  E.~J.~Ludwig, K.~D.~Veal, M.~H.~Wood, A.~Kievsky, S.~Rosati, and
  M.~Viviani, Phys. Rev. C {\bf 63}, 044013 (2001).
\bibitem{Tornow91}W.~Tornow, C.~R.~Howell, M.~Alohali, Z.~P.~Chen,
  P.~D.~Felsher, J.~M.~Hanly, R.~L.~Walter, G.~Mertens, I.~Slaus,
  H.~Witala, and W.~Gl\"ockle, Phys. Lett. B {\bf 257}, 273 (1991).
\bibitem{Witala91}H.~Witala and W.~Gl\"ockle,
  Nucl. Phys. {\bf A528}, 48 (1991); H.~Witala, D.~H\"uber, and W.~Gl\"ockle,
  Phys. Rev. C {\bf 49}, R14 (1994).
\bibitem{Glockle96}W.~Gl\"ockle, H.~Witala, D.~H\"uber, H.~Kamada, and
  J.~Golak, Phys. Rep. {\bf 274}, 107 (1996).
\bibitem{Stoks98}V.~G.~J.~Stoks, Phys. Rev. C {\bf 57}, 445 (1998).
\bibitem{Kievsky97}A.~Kievsky, S.~Rosati, M.~Viviani, C.~R.~Brune, 
H.~J.~Karwowski, E.~J.~Ludwig, and M.~H.~Wood, Phys. Lett. B {\bf 406}, 
292 (1997).
\bibitem{Karwowski99}H.~J.~Karwowski, C.~R.~Brune, W.~H.~Geist,
E.~J.~Ludwig, K.~D.~Veal, M.~H.~Wood, A.~Kievsky, S.~Rosati, M.~Viviani,
and T.~C.~Black, Acta Physica Polonica B {\bf 30}, 1479 (1999).
\bibitem{Wood00}M.~H.~Wood, Ph.D. thesis, University of North Carolina
  at Chapel Hill (2000).
\bibitem{Kievsky00}A.~Kievsky, M.~H.~Wood, C.~R.~Brune, B.~M.~Fisher,
H.~J.~Karwowski, D.~S.~Leonard, E.~J.~Ludwig, S.~Rosati, and
M.~Viviani, Phys. Rev. C
{\bf 63}, 024005 (2001). 
\bibitem{Black95}T.~C.~Black, Ph.D. thesis, University of North Carolina
  at Chapel Hill (1995), Available from University Microfilms, Ann Arbor,
  Michigan, order \#9616149.
\bibitem{Nim} V.~G.~J.~Stoks, R.~A.~M.~Klomp, M.~C.~M.~Rentmeester,
and J.~J.~de~Swart, Phys. Rev. C {\bf 48}, 792 (1993);
  M.~C.~M.~Rentmeester and J.~J.~de Swart,
  preliminary Nijmegen proton-proton PWA97 (private communiation).
\bibitem{Tonsfeldt80}S.~A.~Tonsfeldt, {\it Polarization Phenomena in
    Nuclear Physics}, Part 2, Santa Fe, NM, AIP, 961 (1980);
  S.~A.~Tonsfeldt, Ph.D. thesis, University of North Carolina at
  Chapel Hill, Available from University Microfilms, Ann Arbor,
  Michigan, order \#8022515.
\bibitem{Brune97}C.~R.~Brune, H.~J.~Karwowski, and E.~J.~Ludwig,
Nucl. Instrum. Methods A {\bf 389}, 421 (1997). 
\bibitem{Witala00}H.~Witala, W.~Gl\"ockle, J.~Golak, A.~Nogga, H.~Kamada, 
R.~Skibi\'nski, and J.~Kuros-Zolnierczuk, Phys. Rev. C {\bf 63}, 024007 (2001).
\bibitem{Epelbaum01}E. Epelbaum, H.~Kamada, A.~Nogga, H.~Witala,
W.~Gl\"ockle, and Ulf-G.~Meissner , Phys. Rev Lett. {\bf 86}, 4787 (2001).
\bibitem{Huber98}D.~H\"uber and J.~L.~Friar, Phys. Rev. C {\bf 58},
  674 (1998).
\bibitem{Kievsky99}A.~Kievsky, Phys. Rev. C {\bf 60}, 034001 (1999).
\bibitem{Seyler69}R.~G.~Seyler, Nucl. Phys. {\bf A124}, 253 (1969).
\bibitem{Bevington92}P.~R.~Bevington and D.~K.~Robinson, {\it Data
    Reduction and Error Analysis for the Physical Sciences}, second
  edition, McGraw-Hill, New York (1992).
\bibitem{Sagara94}K.~Sagara, H.~Oguri, S.~Shimizu, K.~Maeda,
H.~Nakamura, T.~Nakashima, and S.~Morinobu, Phys. Rev. C {\bf 50}, 576 (1994).
\bibitem{Sagara00}K.~Sagara, private communication.
\bibitem{Sowinski87}J.~Sowinski, D.~D.~Pun Casavant, and L.~D.~Knutson, Nucl.Phys. {\bf A464}, 223 (1987).
\bibitem{Gruebler83}W.~Gr\"uebler, V.~K\"onig, P.~A.~Schmelzbach,
F.~Sperisen, B.~Jenny, R.~E.~White, F.~Seiler, and H.~W.~Roser,
Nucl. Phys. {\bf A398}, 445 (1983).
\bibitem{Huttel83}E.~Huttel, W.~Arnold, H.~Berg, H.~H.~Krause,
J.~Ulbright, and G.~Clausnitzer, Nucl. Phys. {\bf A406}, 435 (1983).
\bibitem{Shadow01}W.~Shadow and L.~Canton, Phys. Rev. C, in press.
\bibitem{Viviani01}M.~Viviani, A.~Kievsky, S.~Rosati, E.~A.~George,
and L.~D.~Knutson, Phys. Rev. Lett. {\bf 86}, 3739 (2001).
\end{references}
\end{document}